\documentclass{sig-alt-gov2}
\usepackage{times}
\usepackage[latin1]{inputenc}
\usepackage{amsmath}
\usepackage{amsfonts}
\usepackage{amssymb}
\usepackage{verbatim}
\usepackage{algorithm}
\usepackage{algpseudocode}
\usepackage{hyperref}
\usepackage{subfigure}
\usepackage{url}
\usepackage{balance}	
\usepackage{epsfig}
\usepackage{cite}
\algnotext{EndFor}
\algnotext{EndIf}
\algnotext{EndWhile}

\begin{document}

\conferenceinfo{SIGMOD'13,} {June 22--27, 2013, New York, New York, USA.} 
\CopyrightYear{2013} 
\crdata{978-1-4503-2037-5/13/06} 
\clubpenalty=10000 
\widowpenalty = 10000

\title{StreamWorks - A system for Dynamic Graph Search}


\numberofauthors{5} 
\author{
% 1st. author
\alignauthor
Sutanay Choudhury\\
       \affaddr{Pacific Northwest National Laboratory, USA}\\
       %\affaddr{902 Battelle Boulevard}\\
       %\affaddr{Richland, WA, USA 99352}\\
       \email{sutanay.choudhury@pnnl.gov}
% 2nd. author
\alignauthor
Lawrence Holder\\
       %\affaddr{School of Electrical Engineering \& Computer Science}\\
       \affaddr{Washington State University, USA}\\	
       %\affaddr{Pullman, WA, USA 99164-2752}\\
       \email{holder@wsu.edu}
% 3rd. author
\alignauthor
George Chin\\
       \affaddr{Pacific Northwest National Laboratory, USA}\\
       %\affaddr{902 Battelle Boulevard}\\
       %\affaddr{Richland, WA, USA 99352}\\
       \email{george.chin@pnnl.gov}
\and
\alignauthor
Abhik Ray\\
       \affaddr{Washington State University, USA}\\
       %\affaddr{902 Battelle Boulevard}\\
       %\affaddr{Richland, WA, USA 99352}\\
       \email{abhik.ray@wsu.edu}
\alignauthor
Sherman Beus\\
       \affaddr{Pacific Northwest National Laboratory, USA}\\
       %\affaddr{902 Battelle Boulevard}\\
       %\affaddr{Richland, WA, USA 99352}\\
       \email{sherman.beus@pnnl.gov}
%\and  % use '\and' if you need 'another row' of author names
% 4th. author
\alignauthor
John Feo\\
       \affaddr{Pacific Northwest National Laboratory, USA}\\
       %\affaddr{902 Battelle Boulevard}\\
       %\affaddr{Richland, WA, USA 99352}\\
       \email{john.feo@pnnl.gov}
}

\maketitle
\begin{abstract}

Acting on time-critical events by processing ever growing social media, news or cyber data streams is a major technical challenge.  Many of these data sources can be modeled as multi-relational graphs.  Mining and searching for subgraph patterns in a continuous setting requires an efficient approach to incremental graph search.  The goal of our work is to enable real-time search capabilities for graph databases.  This demonstration will present a dynamic graph query system that leverages the structural and semantic characteristics of the underlying multi-relational graph.  
\end{abstract}

%\terms{Algorithms}

\section{Introduction}

Social networks, social media websites and mainstream news media are driving an exponential growth in online content and network traffic. This information barrage presents both a formidable challenge and an opportunity to applications that thrive on situational awareness.  Domains such as emergency response, cyber security, intelligence and finance has many applications that continuously monitor the data stream to look for specific events.  Timeliness of the detection carries paramount importance for such applications.  The applications derive their competitive edge from fast detection as late detection may not have much value due to incurred damage to resources.  Our work is motivated by queries that look for rare events, have a time constraint on the time to discovery and never need a bulk retrieval of historic data due to their monitoring nature.  

The field of relational databases studied the topic of continuous queries to address applications with precisely the above characteristics.  A continuous query (CQ) system is defined as one where a query logically runs continuously over time as opposed to being executed intermittently \cite{Chandrasekaran:2003:TCD:872757.872857, Chen:2000:NSC:342009.335432, Abadi:2003:ANM:950481.950485}.  Many of the prominent news, social media or cyber data streams can be represented as multi-relational graphs.  Following the sprit of CQ systems, our work can be viewed as continuously searching a temporally evolving (henceforth referred as dynamic) graph for graph based patterns representing various events of interest.  

\begin{figure}[]
\centering
\includegraphics[scale=0.95]{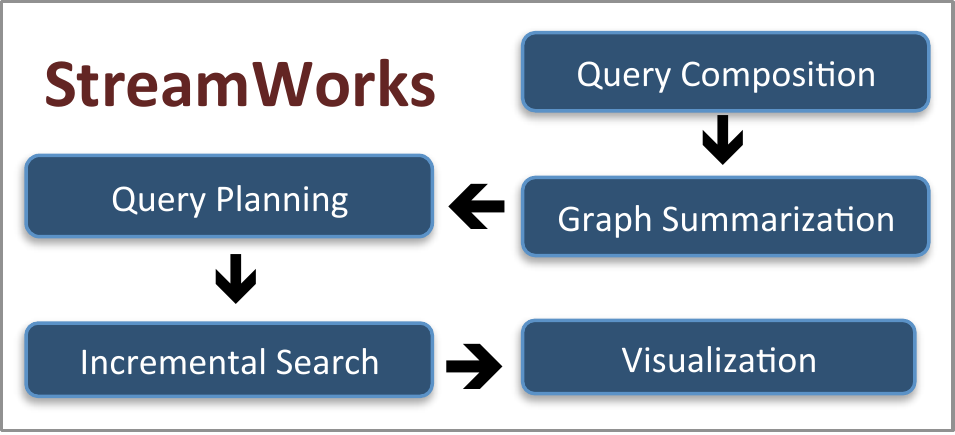}
\caption{Various components for graph mining and search.}
\label{fig:StreamWorks}
\end{figure}

Our proposed demonstration will showcase StreamWorks (Fig. \ref{fig:StreamWorks}) - an analytics framework for dynamic graphs. With StreamWorks, a user can register graph queries to find events as they emerge in the data graph. The novelty of StreamWorks lies in its incremental graph search algorithm based on a query decomposition approach.  The registered queries are decomposed into sub-patterns using a novel data structure called the SJ-Tree \cite{Choudhury:2012:PNNL_Tech_Rep} that systematically tracks the evolution of matches in the underlying graph.  The query decomposition is performed by utilizing statistics and summaries about the data graph such as degree distribution, vertex and edge type distribution and multi-retlational triad distribution.  

\subsection{Demonstration features}

We will present an interface to compose and execute graph queries, and query planning. Further, we will provide visualization of the evolving graph, results from graph queries, and relevant statistics. 

%interface to show the evolving summaries of the dynamic graph.  This
%We will present the following capabilities through an interactive demonstration.\\
%1. A visual interface to show the evolving summaries (such as degree, type and triad distribution) of the dynamic graph.\\
%2. An experimental graph query composition interface and demonstrate query planning features.\\
%3. Register and execute queries on streaming cyber data modeled as a dynamic graph. \\
%4. Visualization of evolving matches and partial match statistics.

%\vfill\eject
\section{Background and Related Work}

\subsection{Problem Statement}
\label{subsec:Problem Statement}

Our theoretical contribution is the development of an incremental subgraph isomorphism algorithm for dynamic graphs \cite{Choudhury:2012:PNNL_Tech_Rep}.  Given a pattern or query graph (henceforth described as query graph) $G_q$ and a larger input graph (henceforth described as the data graph) $G_d$, an isomorphism of $G_q$ in $G_d$ is defined as the matching that involves a one-to-one correspondence between the vertices of a subgraph of $G_d$ and vertices of $G_q$ such that all vertex adjacencies are preserved.  

Every edge in a dynamic graph has a timestamp associated with it and therefore, for any subgraph $g$ of a dynamic graph we can define a time interval $\tau(g)$ which is equal to the interval between the earliest and latest edge belonging to $g$.  Given a dynamic multi-relational graph $G_d$, a query graph $G_q$ and a time window $t_W$, we report whenever a subgraph $g_d$ that is isomorphic to $G_q$ appears in $G_d$ such that $\tau(g_d) < t_W$.  The isomorphic subgraphs are also referred to as \textit{matches} in the subsequent discussions.  If $M(G^k_d)$ is the cumulative set of all matches discovered until time step $k$ and $E_{k+1}$ is the set of edges that arrive at time step $k+1$, we present an algorithm to compute a function $f\left(G_d, G_q, E_{k+1}\right)$ which returns the incremental set of matches that result from updating $G_d$ with $E_{k+1}$ and is equal to $M(G^{k+1}_d) - M(G^k_d)$.

\subsection{Related Work}
\label{sec:Related Work}
Investigation of subgraph isomorphism for dynamic graphs did not receive much attention until recently.  It introduces new algorithmic challenges because we can-not afford to index a dynamic graph frequently enough for applications with real-time constraints.  In fact this is a problem with searches on large static graphs as well \cite{DBLP:journals/pvldb/SunWWSL12}.  There are two alternatives in that direction.  We can search for a pattern repeatedly or we can adopt an incremental approach.  The work by Fan et al. \cite{Fan:2011:IGP:1989323.1989420} presents incremental algorithms for graph pattern matching.  However, their solution to subgraph isomorphism is based on the repeated search strategy.  Chen et al. \cite{Chen:2010:CSP:1850481.1850517} proposed a feature structure called the \textsl{node-neighbor tree} to search multiple graph streams using a vector space approach.  They relax the exact match requirement and require significant pre-processing on the graph stream.  Our work is distinguished by its focus on temporal queries and handling of partial matches as they are tracked over time using a novel data structure.  There are strong parallels between our algorithm and the very recent work by Sun et al. \cite{DBLP:journals/pvldb/SunWWSL12}, where they implement a query-decomposition based algorithm for searching a large static graph in a distributed environment.  Our work is distinguished by the focus on continuous queries that involves maintenance of partial matches as driven by the query decomposition structure, and optimizations for real-time query processing.

\section{Incremental Query Processing}
\label{sec:Incremental Query Processing}
%HOW IS THE DATA STORED?

\subsection{Our Approach}
%\section{A Query Decomposition approach for Continuous Queries}
%\label{sec:Temporal Query Decomposition}

A simplistic approach to solving this problem would be to check, for every edge update, if that edge matches one in the query graph. Once an edge is considered as a matching candidate, the next step is to consider different combinations of matches it can participate in. While intuitively simple, this approach falls prey to combinatorial explosion very quickly.  Our objective is to introduce an approach that guides the search process to look for specific subgraphs of the query graph and follow specific transitions from small to larger matches.  Following are the main intuitions that drive this approach,

1. Instead of looking for a match with the entire graph or just any edge of the query graph, partition the query graph into smaller subgraphs and search for them.\\
2. Track the matches with individual subgraphs and combine them to produce progressively larger matches.\\
3. Define a \textit{join order} in which the individual matching subgraphs will be combined.  Do not look for every possible way to combine the matching subgraphs.

Although the current work is completely focused on temporal queries, the graph decomposition approach is suited for a broader class of applications and queries.  The key aspect here is to search for substructures without incurring too much cost.  Even if some subgraphs of the query graph are matched in the data, we will not attempt to assemble the matches together without following the join order.  Thus, if there are substructures that are too frequent, joining them and producing larger partial matches will be too expensive without a stronger guarantee of finding a complete match.  On the other hand, if there is a substructure in the query that is rare or indicates high selectivity, we should start assembling the partial matches together only after that substructure is matched.  
\begin{comment}
Thus, for query graphs that have different substructures with varying frequency or selectivity, the problem of assembling partial matches is equivalent to a join order optimization problem \cite{Swami:1989:OLJ:66926.66961}.   \end{comment}
%\vspace{12 mm}
\begin{figure}[]
\includegraphics[scale=0.42,height=75mm]{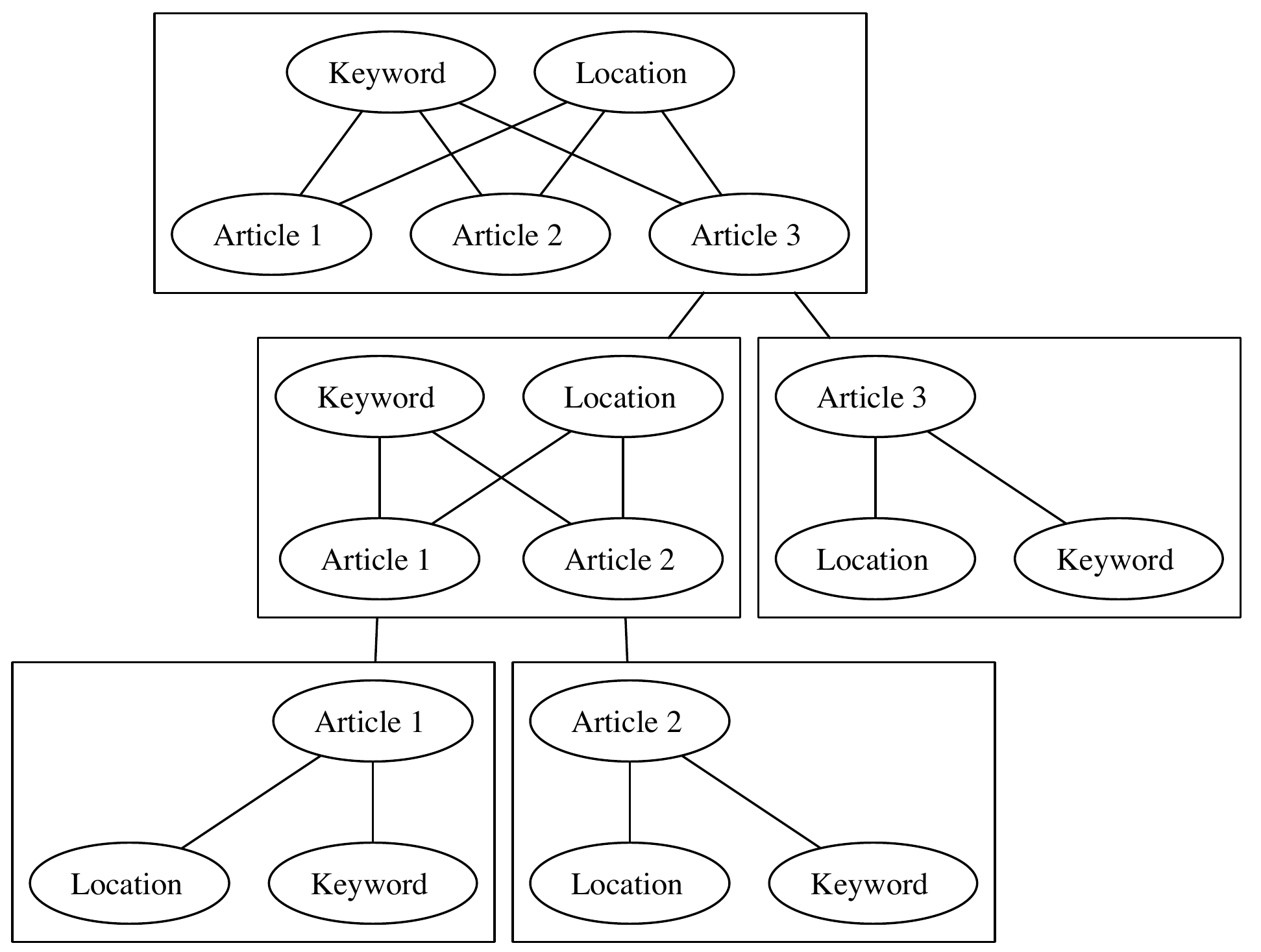}
\caption{Illustration of query decomposition in SJ-Tree.  The graph shown in the root node represents a query to find three articles or posts with a common keyword and location.}
\label{fig:join_tree}
\end{figure}

\subsection{Subgraph Join Tree (SJ-Tree)}
\label{subsubsec:join_tree_properties}
We introduce a tree structure called \emph{Subgraph Join Tree (SJ-Tree)} that supports the above intuitions for implementing a search and join order based on selectivity of substructures of the query graph.  Fig. \ref{fig:join_tree} shows an example decomposition of a query graph.

\textsc{Definition 4.1.1 } A SJ-Tree $T$ is defined as a binary tree comprised of the node set $N_T$.  Each $n \in N_{T}$ corresponds to a subgraph of the query graph $G_q$.   Let's assume $V_{SG}$ is the set of corresponding subgraphs and $|V_{SG}| = |N_T|$.  Additional properties of the SJ-Tree are defined below.

\textsc{Property 1.} The subgraph corresponding to the root of the SJ-Tree is isomorphic to the query graph.  Thus, for $n_r = root\lbrace T\rbrace$, $V_{SG}\lbrace n_r \rbrace \equiv G_q$.

\textsc{Property 2.} The subgraph corresponding to any internal node of $T$ is isomorphic to the output of the join operation between the subgraphs corresponding to its children.  If $n_l$ and $n_r$ are the left and right child of $n$, then $V_{SG}\lbrace n \rbrace = V_{SG}\lbrace n_l \rbrace \Join V_{SG}\lbrace n_r \rbrace$.  Given two graphs $G_1 = (V_1, E_1)$ and $G_2 = (V_2, E_2)$, the join operation is defined as $G_3 = G_1 \Join G_2$, such that $G_3 = (V_3, E_3)$ where $V_3 = V_1 \cup V_2$ and $E_3 = E_1 \cup E_2$.

\textsc{Property 3.} Each node in the SJ-Tree maintains a set of matching subgraphs.  We define a function $matches(n)$ that for any node $n \in N_T$, returns a set of subgraphs of the data graph.  If $M = matches(n)$, then $\forall G_m \in M$,  $G_m \equiv V_{SG}\lbrace n \rbrace$.

\textsc{Property 4.} Each internal node $n$ in the SJ-Tree maintains a subgraph, CUT-SUBGRAPH($n$) that 
equals the \textit{intersection} of the query subgraphs of its child nodes.  

\section{System Overview}

\subsection{Query Planning}

With the subgraph join-tree data structure in mind, the next task is to automatically decompose a query graph $G_q$ and create a subgraph join tree based on the decomposition. Broadly our aim is to decompose the query graph into a number of smaller graphs, which we refer to as \textsl{search primitives}, and perform local searches for these primitives.  We use the term \textit{local search} to refer to a subgraph search performed in the neighborhood of an edge in the data graph for a \textit{small query subgraph}.  The primitives are restricted to small and "selective"  query subgraphs to keep the local search efficient.  An important goal of the decomposition process is to push the most selective subgraph at the lowest level in the subgraph join-tree to reduce the number of partial matches.

\subsection{Query Execution}

Our proposed subgraph matching algorithm contains two primary tasks.  First, for every incoming edge we perform a local search to detect a match with the smallest subgraphs associated with the leaves of the SJ-Tree.  When a match is found with the subgraph corresponding to the leaf node of the SJ-Tree, we initialize a match structure and insert it into the collection maintained at that leaf node.  Upon insertion of a match into a leaf node we check to see if it can be combined with any matches that are contained in the collection maintained at its sibling node.  A successful combination of matching subgraphs between the leaf or intermediate node and its sibling node leads to the insertion of a larger match at the respective parent node.  This process is repeated as long as larger matching subgraphs can be produced by moving up in the SJ-Tree.  A complete match is found when two matches belonging to the children of the root node are combined successfully. 

\subsection{Summarization}

Summarization involves collecting summary statistics about the data graph to use for query planning.  We collect three different types of information 1) degree distribution 2) distribution of vertex and edge types, 3) the frequency distribution of multi-relational triad structures.  Incorporation of triad statistics into the query decomposition process is a work in progress at the time of this writing.  Continuously collecting the statistics information from the data stream and updating the query decomposition and search strategy remains an area for future work.

\section{Target Applications}

We focus on two major application domains: cyber-security and news/social media monitoring.  The following subsections present a quick snapshot of some motivating queries.  

\subsection{Cyber-Security}
\begin{figure}[]
\centering
\includegraphics[scale=0.3]{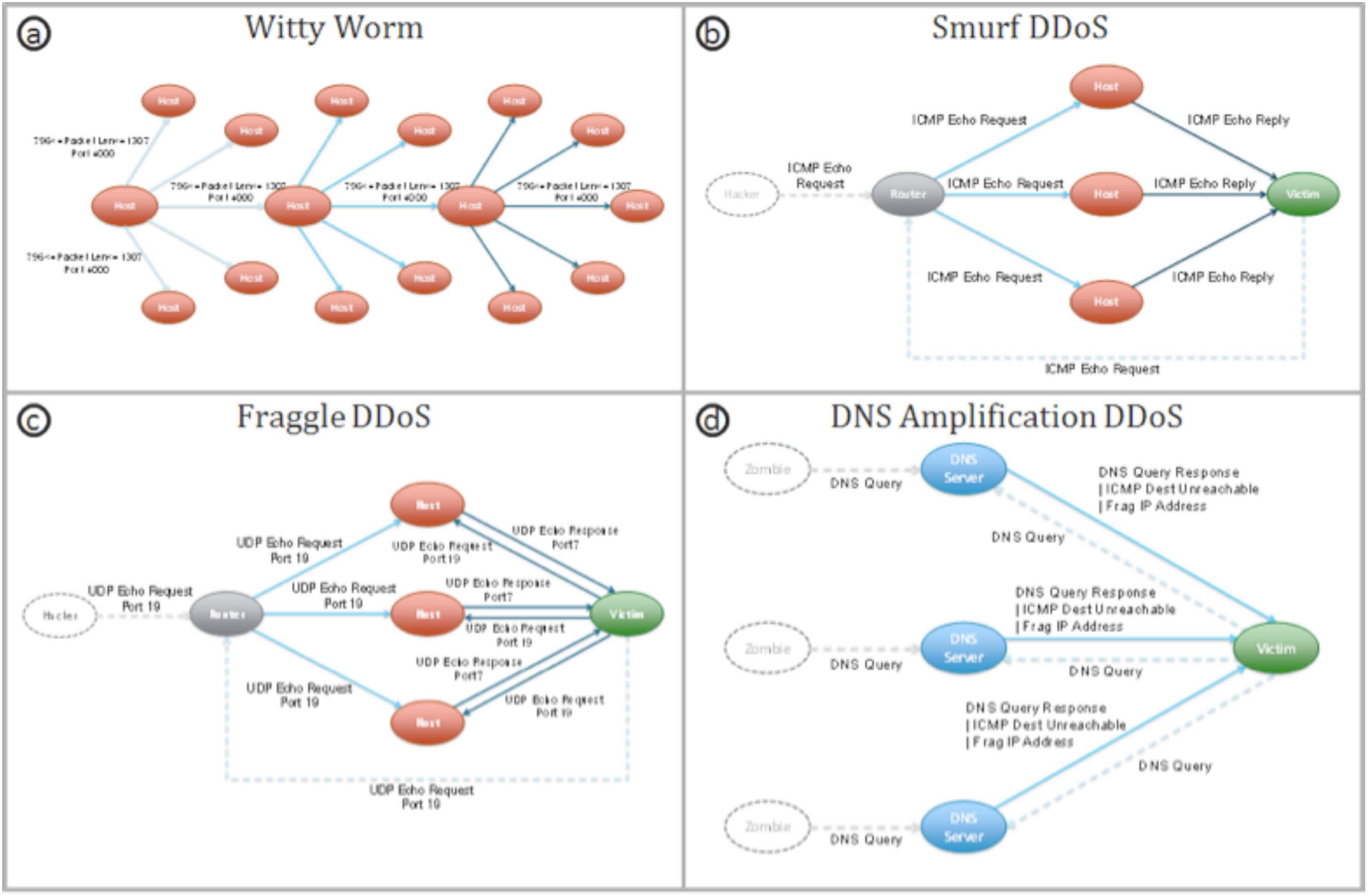}
\caption{Examples queries to detect cyber attacks.}
\label{fig:attack_examples}
\end{figure}

A cyber system is naturally described as a graph with physical machines, IP addresses, users, and software services as entities (vertices).  The relationships between these entities such as communication between machines, association of a physical machine and an IP address, login of a user on a machine etc are modeled as edges in the graph.   From a security perspective, updates to this dynamic graph can be constantly monitored to detect events such as worm spread, virus attack, denial-of-service attack etc..  We construct graph-based representation of these events (Fig. \ref{fig:attack_examples}) and query the data graph to detect occurrences of malicious events.  

%\vfill\eject
\subsection{News and Social Media}
Various online news or social media data sources can be represented as multi-relational graphs.  Entities such as articles, events, people, location, organizations and keywords can be represented as vertices in the graph.  Next, graph based queries can be executed to detect the occurrence of various events in the news stream.  Fig. shows some example queries and Fig. \ref{fig:geoloc_queries} shows a map-based visualization of the a series of queries executed on New York Times data \footnote[1]{http://data.nytimes.com}.  

\section {Demonstration Setup}
\subsection{Setup}
\textbf{Dataset}:  We will demonstrate the query capabilities on internet traffic data obtained from www.caida.org.  The number of records in these datasets typically varies between 50-100 million/hour.

\textbf{Software/Hardware} The queries will be executed on a 48-core shared memory system running Linux 2.6.18 and comprising 2.3 GHz AMD Opteron 6176 SE processors and 256 GB memory.  Each system node has 32 GB memory attached to it.  The graph query engine is implemented in C++.  

\subsection{User Interface}

\begin{figure}[h!]
\centering
\includegraphics[scale=0.33]{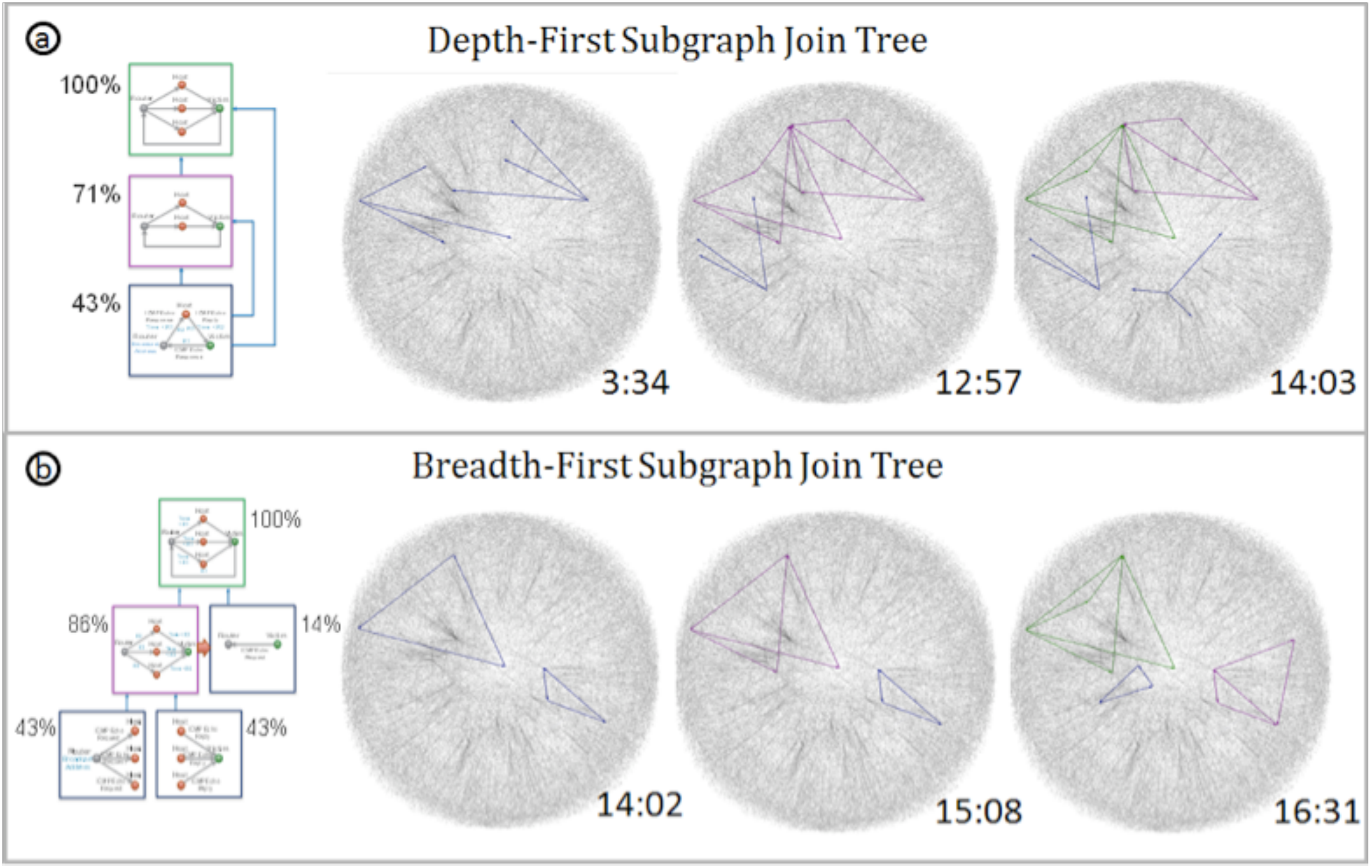}
\caption{Emerging matches for Smurf DDoS subgraph patterns in a dynamic computer network using different query plans.}
\label{fig:cyber_match_evolution}
\end{figure}

There are three major focus areas for visualization and UI design. 
\begin{itemize}
\item  Our primary target audience includes journalists, emergency responders, intelligence professionals who are not expected to use StreamWorks using an API.  Fig.  \ref{fig:UI} shows an experimental user interface for visual query composition.  The user interface will retrieve metadata information such as vertex and edge types and their attributes to assist in drawing a query graph.  
\begin{figure}[]
\centering
\includegraphics[scale=0.25]{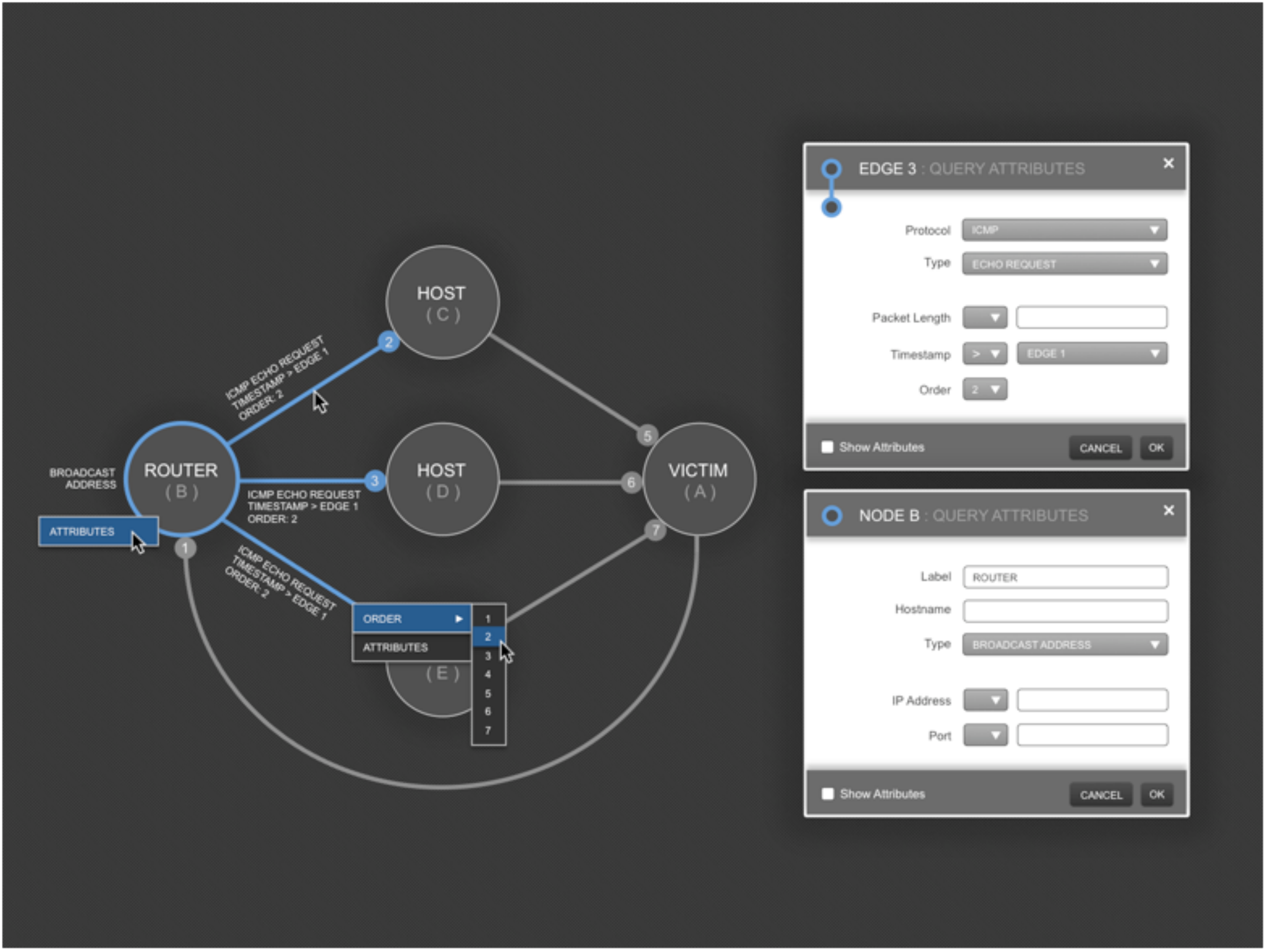}
\caption{Prototype of an interface for visual graph query composition.}
\label{fig:UI}
\end{figure}
\item  The query graphs are a representation of events of interest; hence, we are developing map (Fig. \ref{fig:geoloc_queries})  and tabular views (Fig. \ref{fig:cyber_UI}) that show occurrence of events in a geospatial and temporal context.  The goal is to keep the underlying graph representation transparent to the user.   Query results from any graph with location information available as a vertex attribute can be displayed on the map view.  
\begin{figure}[h]
\centering
\includegraphics[scale=0.21]{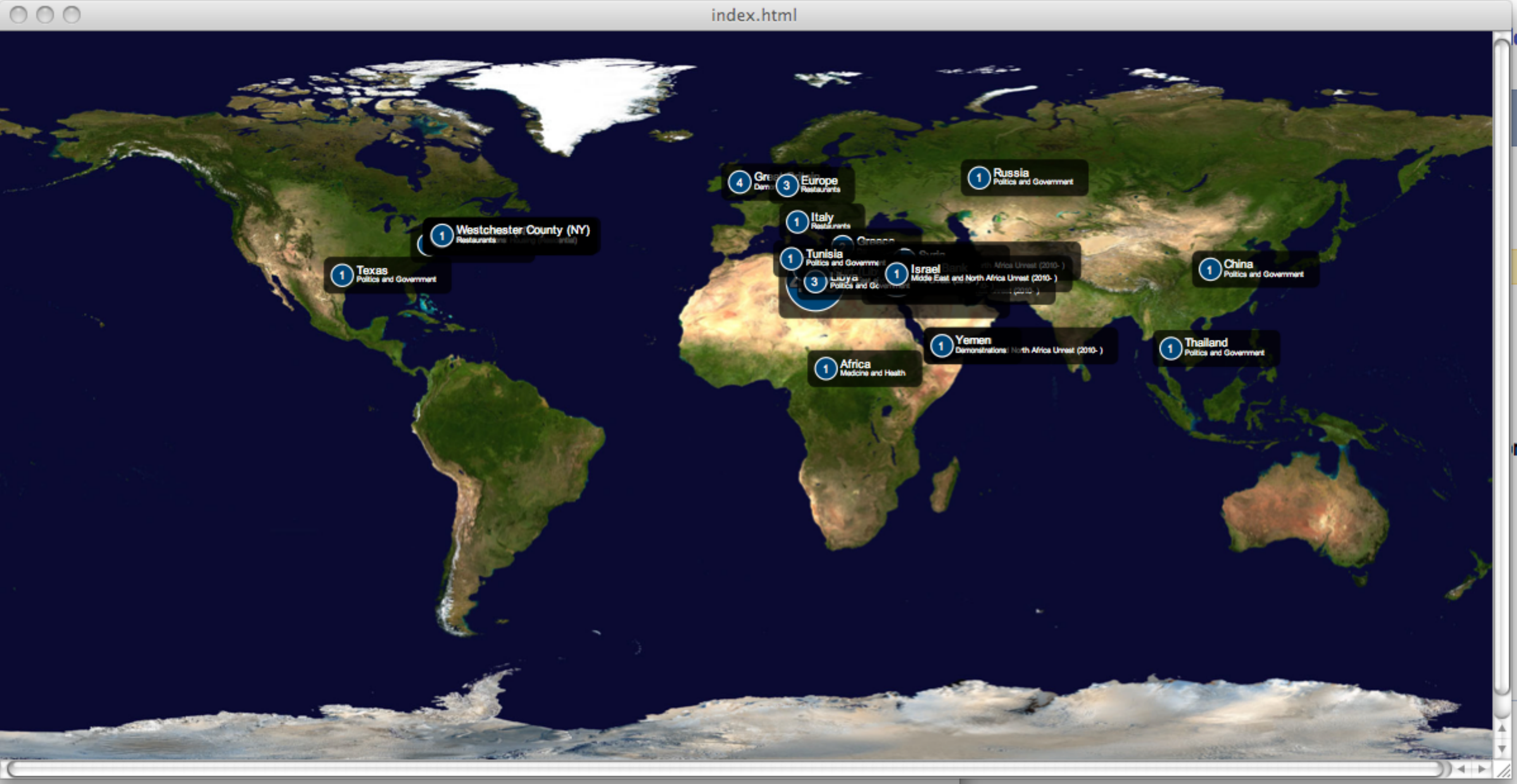}
\caption{A visualization of the output from a collection of graph queries.  The queries are similar to Fig. \ref{fig:join_tree}.  Each query graph specifies a label (such as "politics", "accident" etc.) on the keyword vertex to indicate the event of interest.}
\label{fig:geoloc_queries}
\end{figure}
\begin{figure}[]
\centering
\includegraphics[scale=0.25]{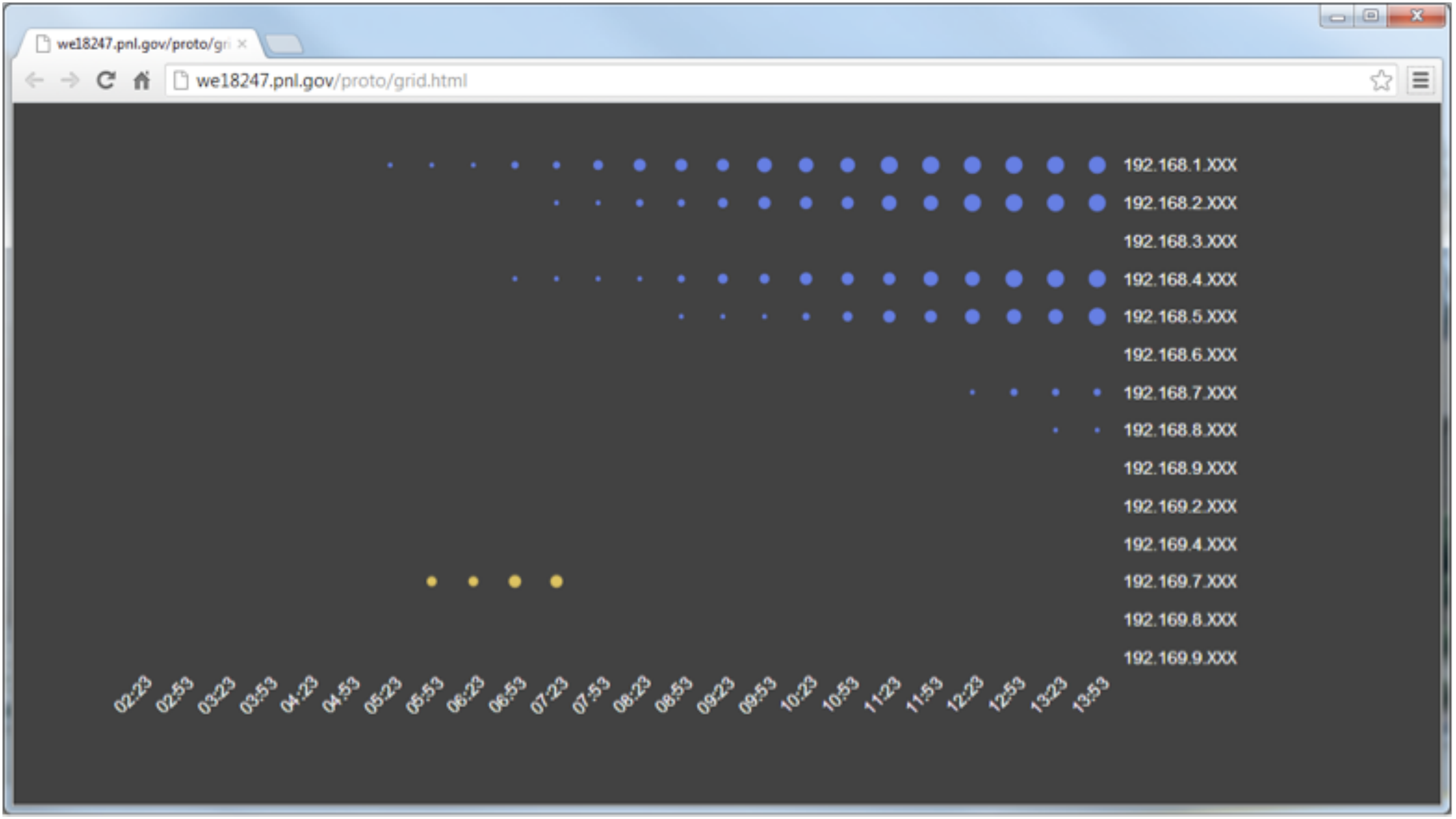}
\caption{Grid-based visualization showing cascading effect of a Smurf DDoS attack across subnetworks (blue dots).}
\label{fig:cyber_UI}
\end{figure}
\item Graph-based visualization of the results from subgraph queries is critical for developers and API users.  Therefore, we are adapting and applying the Gephi graph visualization and manipulation software \cite{sfw:gephi} to render snapshots of the data graph and encode the partial and complete matches.  This is also useful to observe the choice of different query decomposition strategies. To illustrate, Fig. \ref{fig:cyber_match_evolution} shows snapshots of emerging subgraph patterns in a computer network that are identified and tracked using different SJ-Tree structures.  The percentages show the fraction of query graph being matched as measured by the number of edges.  Each SJ-Tree is shown next to its associated emerging subgraph pattern snapshots. The colors of the subgraph patterns in the snapshots correspond to particular partitions in the associated SJ-Tree to indicate the level or degree of partial matching to the query graph.
\end{itemize} 
%The matching scores or percentages are shown for each SJT.

%We used the CAIDA dataset \cite{data:caida}, which we augmented to superimpose a Smurf DDoS attack. To demonstrate the function of the SJTs, we display each SJT next to its associated emerging subgraph pattern snapshots. The colors of the subgraph patterns in the snapshots correspond to particular partitions in the associated SJT to indicate the level or degree of partial matching to the query graph. The matching scores or percentages are shown for each SJT.

\section{Acknowledgments}
Presented research is based on work funded under the CASS-MT project at Pacific Northwest National Laboratory, which is operated by Battelle Memorial Institute.

\bibliographystyle{abbrv}
\bibliography{citations_sutanay}  % sigproc.bib is the name of the Bibliography in this case

\balancecolumns
% That's all folks!
\end{document}